\documentclass[twocolumn,prb,aps,superscriptaddress,showpacs,10pt
]{revtex4-1}
\usepackage{amsmath}
\usepackage{amssymb}
\usepackage{graphicx}
\usepackage{dcolumn}
\usepackage{graphicx}
\usepackage{dcolumn}
\usepackage{bm}

\def\be{\begin{equation}}  
\def\ee{\end{equation}}  
\def\ba{\begin{eqnarray}}  
\def\ea{\end{eqnarray}}  
\def\bc{\begin{center}}  
\def\ec{\end{center}}

\begin{document}

\title{Giant microwave-induced $B$-periodic magnetoresistance oscillations in a two-dimensional electron gas with a bridged-gate tunnel point contact}

\author{A. D. Levin}
\affiliation{Instituto de F\'{\i}sica da Universidade de S\~ao
Paulo, 135960-170, S\~ao Paulo, SP, Brazil}

\author{S. A. Mikhailov}
\email[Email: ]{sergey.mikhailov@physik.uni-augsburg.de}
\affiliation{Institute of Physics, University of Augsburg, D-86135 Augsburg, Germany}

\author{G. M. Gusev}
\email[Email: ]{gusev@if.usp.br}
\affiliation{Instituto de F\'{\i}sica da Universidade de S\~ao
Paulo, 135960-170, S\~ao Paulo, SP, Brazil}

\author{Z. D. Kvon}
\affiliation{Institute of Semiconductor Physics, Novosibirsk, 630090 Russia}
\affiliation{Novosibirsk State University, Novosibirsk, 630090 Russia}

\author{E. E. Rodyakina}
\affiliation{Institute of Semiconductor Physics, Novosibirsk, 630090 Russia}

\author{A. V. Latyshev}
\affiliation{Institute of Semiconductor Physics, Novosibirsk, 630090 Russia}
\affiliation{Novosibirsk State University, Novosibirsk, 630090 Russia}

\date{\today}
\begin{abstract}
We have studied the magnetoresistance of the quantum point contact fabricated on the high mobility two-dimensional electron gas (2DEG) exposed to microwave irradiation. The resistance reveals giant $B$-periodic oscillations with the relative amplitude $\Delta R/R$ of up to $700$\% resulting from the propagation and interference of the edge magnetoplasmons (EMPs) in the sample. This giant photoconductance is attributed to the considerably large local electron density modulation in the vicinity of the point contact. We have also analyzed the oscillation periods $\Delta B$ of the resistance oscillations and, comparing the data with the EMP theory, extracted the EMP interference length $L$. We have found that the length $L$ substantially exceeds the distance between the contact leads but rather corresponds to the distance between metallic contact pads measured along the edge of the 2DEG. This resolves existing controversy in the literature and should help to properly design highly sensitive microwave and terahertz spectrometers based on the discussed effect. 
\end{abstract}

\pacs{
73.23.-b
}

\maketitle

A number of interesting magnetotransport phenomena were discovered in recent years in a two-dimensional electron gas (2DEG) placed in a perpendicular magnetic field $B$ and exposed to microwave (MW) irradiation. In low magnetic fields, corresponding to the condition $\omega_{c}\lesssim \omega$, the so called microwave-induced resistance oscillations (MIRO), with zero-resistance states, were observed \cite{Zudov01,Ye01,Mani02,Zudov03}; here $\omega_{c}=eB/m^{\ast}c$ and $\omega=2\pi f$ are the cyclotron and microwave frequencies and $m^{\ast}$ is the electron effective mass. These oscillations of the diagonal resistance $R$ are periodic in $1/B$ and are especially pronounced at low temperatures $T\simeq 1$ K in samples with a high electron mobility. 

In the opposite regime of higher magnetic fields, $\omega_{c} \gtrsim \omega$, another type of magnetoresistance oscillations was discovered \cite{Kukushkin04a}. These oscillations are periodic in $B$ and were explained by the excitation and interference of edge magnetoplasmons\cite{Volkov88Eng} (EMPs) in the sample. The incident microwave radiation excites oscillating (dipole) electric field at the boundaries between the 2DEG and metallic contacts. These dipole fields generate plasma waves -- EMPs -- propagating along the sample edge between the contacts on which the magnetoresistance $R$ is measured. Due to the interference of EMPs, generated by different contacts, $R$ oscillates as a function of the parameter $qL$, where $q(\omega,B)$ is the EMP wave-vector and $L$ is the inter-contact distance \cite{Kukushkin04a}. In strong magnetic fields $q(\omega,B)$ is proportional to $\omega B/n_s$, where
$n_{s}$ is the electron density \cite{Volkov88Eng,Mikhailov06b}. This leads to the $B$-periodic oscillations with the period $\Delta B\propto n_{s}/\omega L$. This theoretically predicted dependence was confirmed by the experiments \cite{Kukushkin04a,Kukushkin05a}. EMPs have a much smaller damping, as compared to the bulk magnetoplasmons,\cite{Volkov88Eng,Mikhailov06b} and, in contrast to the MIRO-ZRS effect, the EMP-related oscillations do not require low temperatures and samples with the very high electron mobility: they were observed at least up to $T\simeq 80$ K \cite{Kukushkin05a}. These EMP properties make the considered phenomenon especially promising for the creation of miniature microwave frequency-sensitive detectors and spectrometers. 

\begin{figure}[ht!]
\includegraphics[width=0.49\columnwidth]{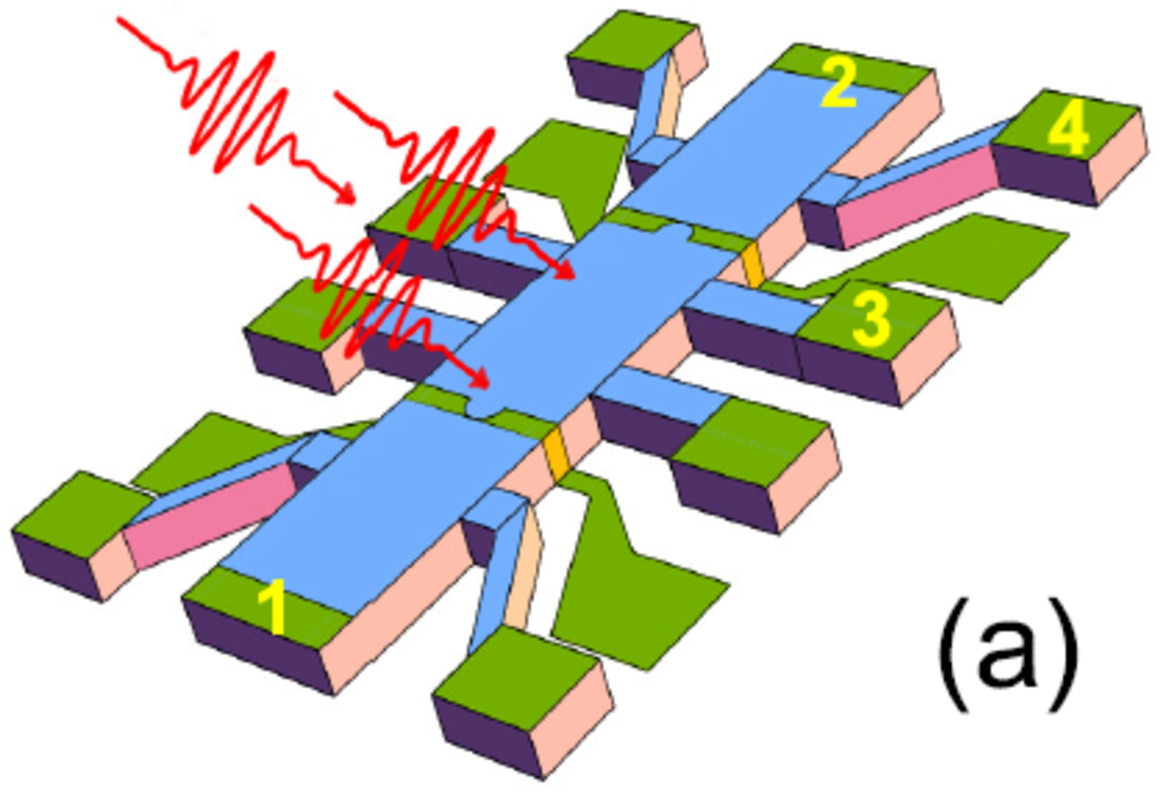}\hspace{0.02\columnwidth}
\includegraphics[width=0.39\columnwidth]{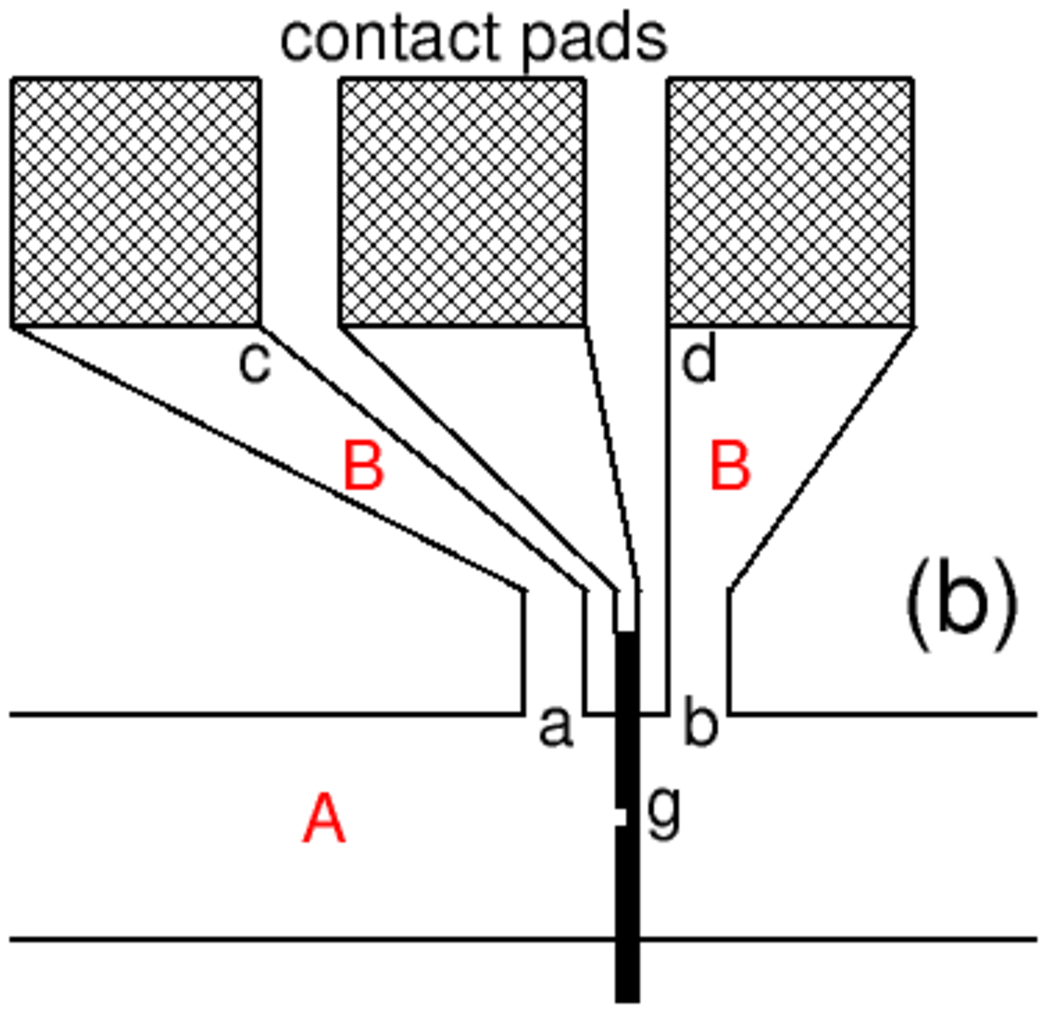}\\
\includegraphics[width=0.9\columnwidth]{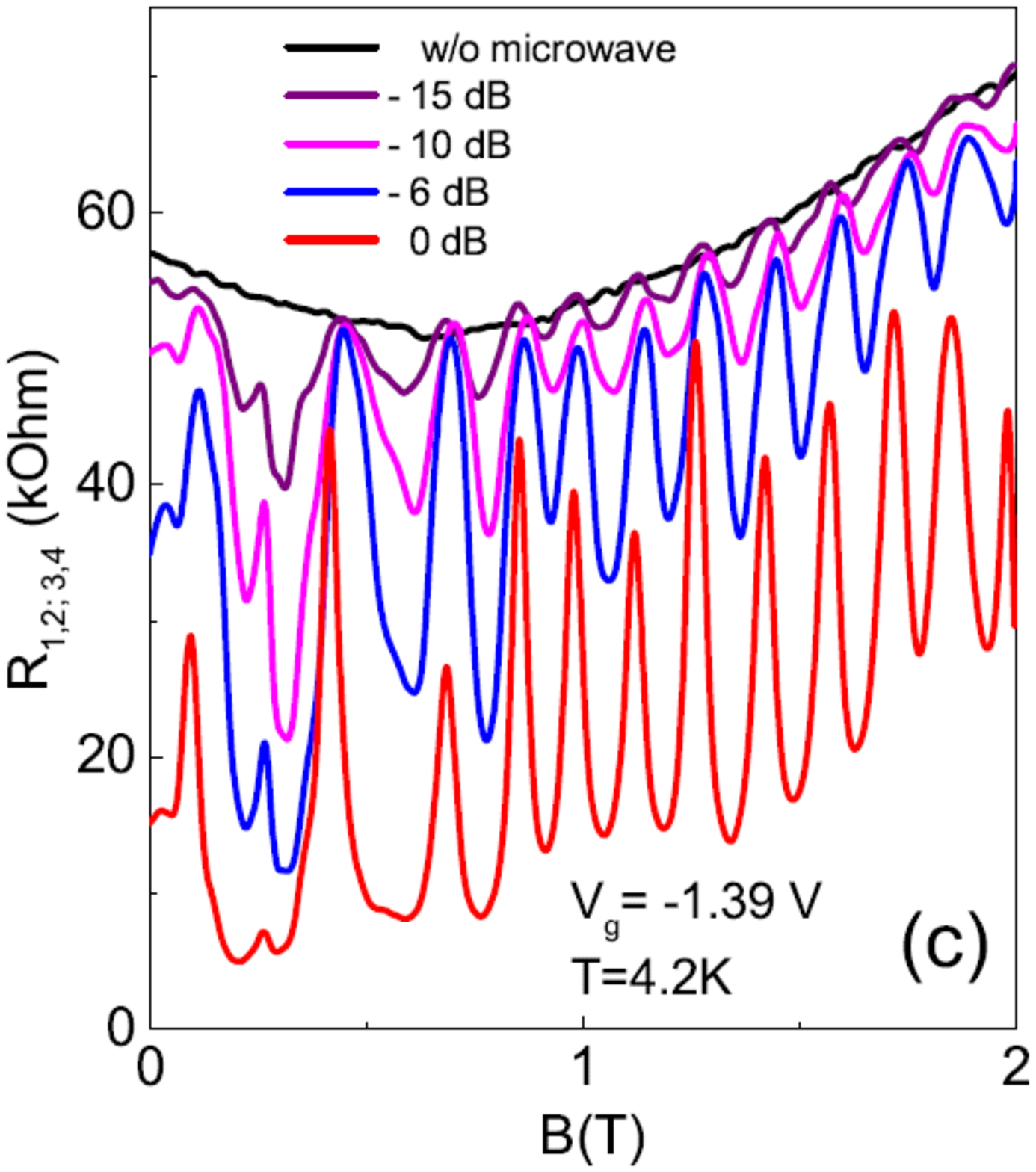}
\caption{\label{fig.1}(Color online) (a) A set-up of the device; the current flows from the contact 1 to the contact 2, the voltage is measured between the contacts 3 and 4; a magnetic field $B$ is perpendicular to the 2DEG plane. (b) Geometry of the metallic contact pads (hatched areas), contact leads (areas B) and of the 2DEG (area A); the black vertical element labeled as `g' is the bridge-gate QPC. (c) The longitudinal resistance $R_{1,2;3,4}\equiv R$ at the microwave frequency $f=156$ GHz, temperature $T= 4.2$ K, the gate voltage $V_{g}=-1.39$ V and different values of the microwave power density.}
\end{figure}

The frequency and density dependence of the oscillation period $\Delta B\propto n_{s}/\omega $, experimentally observed in the first papers \cite{Kukushkin04a,Kukushkin05a}, was confirmed in a later publication by Stone et al. \cite{Stone07}. However, in contrast to \cite{Kukushkin04a,Kukushkin05a}, they found that $\Delta B$ does not depend on the distance $L$ between the contacts. This contradiction remains unexplained so far; actually, it raises the question which length should be understood under $L$ in the discussed phenomenon. Notice that both in Refs. \cite{Kukushkin04a,Kukushkin05a} and Ref. \cite{Stone07} $L$ was assumed to be equal to the distance $L_{ab}$ between the points $a$ and $b$ in Figure \ref{fig.1}(b), i.e. the points where the contact leads (made out of the 2DEG) touch the 2DEG channel. 

In this paper we investigate the EMP-related magnetoresistance oscillations in GaAs/AlGaAs quantum-well structures. Our work differs from the previous publications \cite{Kukushkin04a,Kukushkin05a,Stone07} in three essential aspects. First, the distance $L_{ab}$ in our samples ($\simeq 60$ $\mu$m) is almost two orders of magnitude shorter than in Refs. \cite{Kukushkin04a,Kukushkin05a,Stone07}. If the oscillation period was proportional to $1/L_{ab}$, we would observe more than ten times larger period $\Delta B$ than in Refs. \cite{Kukushkin04a,Kukushkin05a,Stone07}. But in our experiment $\Delta B\simeq 0.1-0.2$ T, which is close to the oscillation periods in \cite{Kukushkin04a,Kukushkin05a,Stone07}. Second, we have performed a more careful, quantitative comparison of our experimental data with the theory \cite{Volkov88Eng} and extracted the value of $L$ from the experimental data. It turns out to be of 1 mm scale which is much larger than $L_{ab}$ but is of order of the distance between the metallic contact pads measured along the edge of the 2DEG. Third, our samples were covered by a thin metallic gate forming a bridged-gate quantum point contact (QPC) \cite{Levin15}. Recently, in Ref. \cite{Levin15}, we have shown (at $B=0$) that the Hall bars with the bridged-gate QPC demonstrate a two-three orders of magnitude higher sensitivity to the microwave irradiation than the structures with a traditional split-gate QPC. In the present work ($B\neq 0$, bridged-gate) we have found a much higher oscillation amplitude of the $B$-periodic oscillations ($\Delta R/R \simeq 7$) than in the ungated structures \cite{Kukushkin04a,Kukushkin05a,Stone07} or in structures with a split-gate QPC \cite{Dorozhkin05} ($\Delta R/R < 1$). Our results lead to a better understanding of the physics of the EMP-related magnetoresistance oscillations and pave the way to the creation of more efficient frequency sensitive detectors of microwave and terahertz radiation.

Our specimens are narrow ($14$ nm)  quantum wells with the electron density $n_{s} \simeq 10^{12}$ cm$^{-2}$ and the mobility of $\mu\simeq 1.3\times 10^{6}$ cm$^{2}$/V s at $4.2-1.4$ K. We measure the resistance on a conventional Hall bar patterned structure, which is designed for
multi-terminal measurements. The sample consists of three $20$ $\mu$m wide consecutive segments of different length ($60$, $20$, $60$ $\mu$m), and eight voltage probes, see Figure \ref{fig.1}(a). High quality Ohmic contacts to a 2DEG  are made using AuGeNi  metalization followed by an annealing. Two thin metallic gates (bridged gate QPCs) are sputtered on the central parts of the left and right segments of the Hall bar. The middle segment was used to measure the resistance of the unpatterned 2D electron gas. Four devices were studied and the similar results were obtained. The measurements were carried out in a VTI cryostat with a waveguide to deliver MW irradiation (the frequency range 110 to 170 GHz, the power density $\sim 1-10$ mW/cm$^2$) down to the sample and by using a conventional lock-in technique to measure the longitudinal resistance $R=R_{1,2;3,4}$.

\begin{figure}[ht!]
\includegraphics[width=\columnwidth]{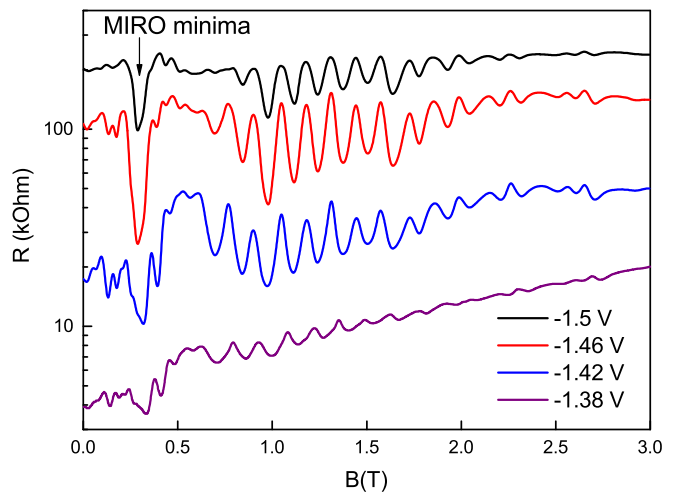}
\caption{\label{fig.2}(Color online) The longitudinal resistance for different gate voltages $V_{g}$ at the MW frequency $f=156$ GHz and temperature $T=4.2$ K. The MW power attenuation is $0$ dB. }
\end{figure}

Figure \ref{fig.1}(c) shows the magnetoresistance of the QPC at the temperature $T=4.2$ K, the frequency $f=156$ GHz and several values of the microwave power attenuation factor. The microwave irradiation leads to the strong suppression of the resistance at zero magnetic field \cite {Levin15}. At $B\gtrsim 0.7$ T (corresponds to $\omega_c\gtrsim \omega$) the resistance reveals the large $B$-periodic magnetoresistance oscillations. Figure \ref{fig.2} illustrates the magnetoresistance at different (fixed) gate voltages varying from the open-contact regime ($R_{0}\lesssim h/e^{2}\simeq 25.8$ k$\Omega$) to the close-contact regime ($R_{0}\gg h/e^{2}$); here $R_{0}$ is the QPC resistance at $B=0$. The oscillation frequency shows no change with $V_{g}$, while the resistance $R_{0}$ increases by several orders of magnitude. The amplitudes of the $B$-periodic oscillations are vanishingly small in the unbiased ($V_{g}=0$), as well as in the unpatterned (no gate), structures. When the absolute value of $V_{g}$ grows, the oscillation amplitudes first increase (blue and red curves in Figure \ref{fig.2}) but then decrease again (black curve) since at very large negative gate voltages the contacts 3 and 4 becomes decoupled. 

\begin{figure}[ht!]
\includegraphics[width=0.9\columnwidth]{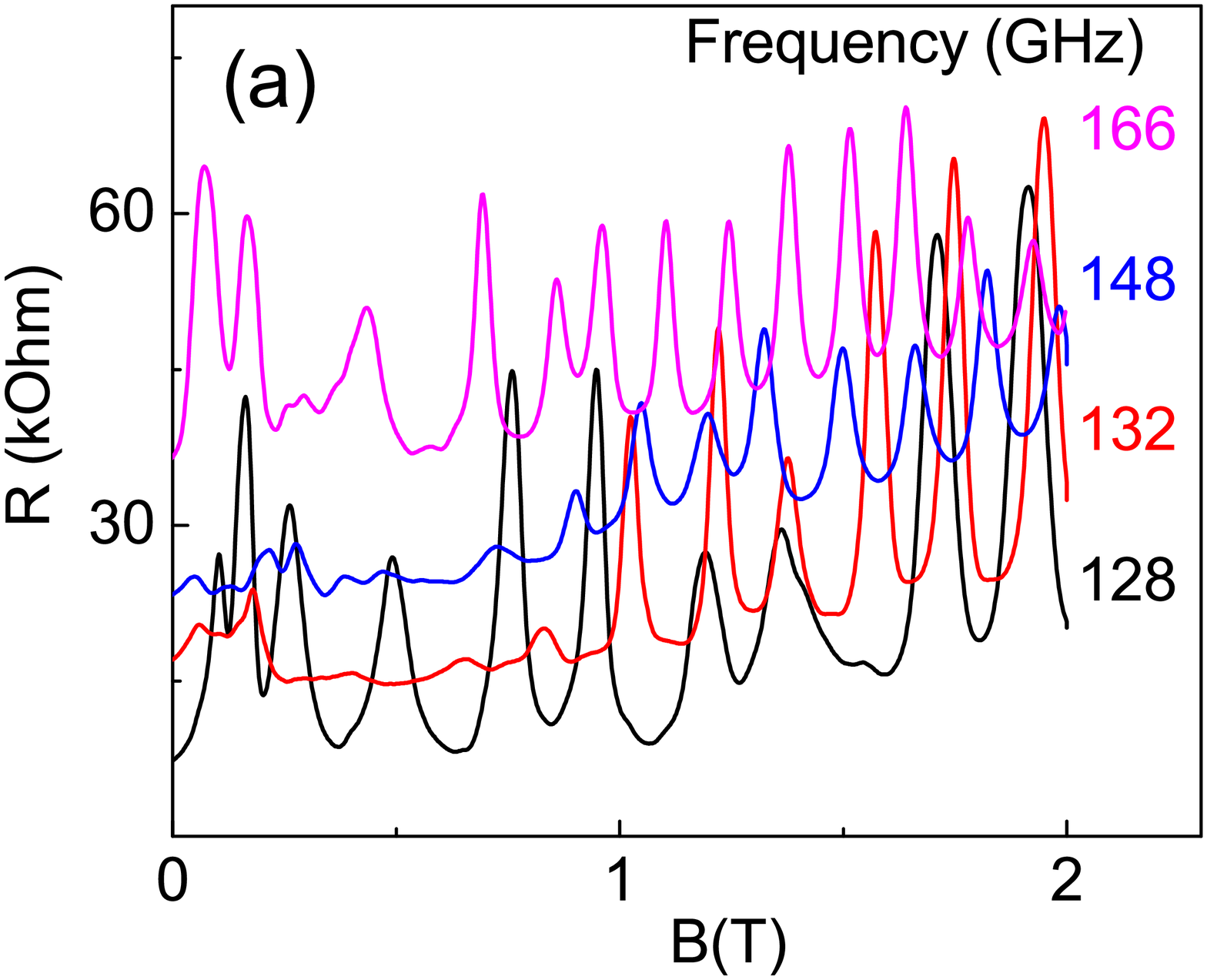}
\includegraphics[width=0.9\columnwidth]{Fig3b.eps}
\caption{\label{fig.3}(Color online) (a) The magnetoresistance $R(B)$ for the frequencies $f=128$, $132$, $148$, and $166$ GHz, temperature $T=4.2$ K, MW power attenuation 0 dB and the gate voltage $V_{g}=-1.39$ V. (b) The interference parameter $qL/2\pi$ calculated from Eqs. (\ref{de})--(\ref{eps}) (continuous curves) and the oscillation index (symbols) as a function of $B$. The fitting length $L$ is chosen in such a way that the theoretical curves coincided with the experimental positions of the oscillation maxima at integer values of $qL/2\pi$; for details of the fitting procedure see the text.}
\end{figure}

Let us analyze the oscillating behavior of $R(B)$ quantitatively. Figure \ref{fig.3}(a) shows the QPC resistance $R$ at four chosen frequencies from the interval $128 - 166$ GHz. At $B\gtrsim 0.7$ T one sees about ten maxima of the $R(B)$ dependence, with the oscillation periods growing with the decreasing frequency. We assume \cite{Kukushkin04a} that the oscillations are caused by the interference of EMPs excited from two points at the sample boundary separated by a length $L$ (e.g. by the distance between some contacts) and the maxima of $R(B)$ are the case when $qL/2\pi$ is integer. Then the length $L$ can be found as follows. 

\begin{figure}[ht!]
\includegraphics[width=\columnwidth]{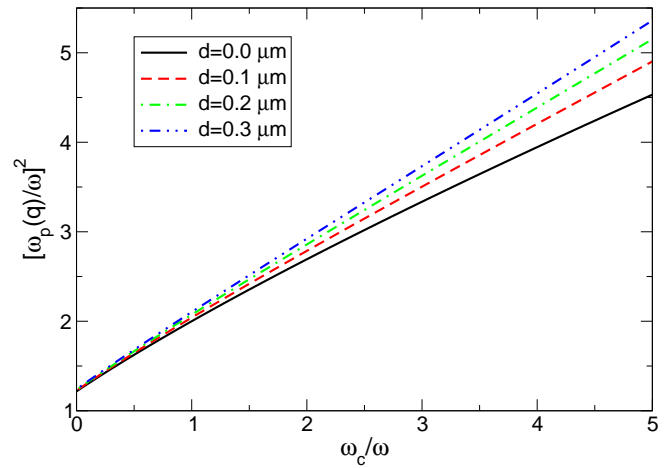}
\caption{\label{fig.4}(Color online) The calculated EMP dispersion relation (\ref{de})--(\ref{eps}) plotted as the dimensionless wave-vector $[\omega_p(q)/\omega]^2=(2\pi n_se^2/m^\ast\bar\kappa\omega^2)q$ versus dimensionless magnetic field $\omega_c/\omega=(e/m^\ast c\omega)B$. The three finite-$d$ curves are plotted for $n_s=10^{12}$ cm$^{-2}$, $f=156$ GHz and parameters of GaAs ($\kappa=12.8$, $m^\ast=0.067$).}
\end{figure}

The spectrum of EMPs running along the boundary of a semi-infinite 2DEG is determined by the dispersion relation \cite{Volkov88Eng}
\be 
\frac{i|q|\sigma_{xx}(\omega)}{q\sigma_{yx}(\omega)}- \tanh\left\{\int_0^{\pi/2}\ln\left[\epsilon\left(\frac{|q|}{\sin t},\omega\right)\right]\frac{dt}\pi\right\}=0,\label{de}
\ee
where $q$ and $\omega$ are the EMP wavevector and frequency, $\sigma_{\alpha\beta}(\omega)$ is the 2DEG Drude conductivity tensor and 
\be 
\epsilon(q,\omega)=1+\frac{4\pi i \sigma_{xx}(\omega)q}
{\omega\kappa\left[\frac{\kappa\tanh(qd)+1}{\kappa+\tanh(qd)}+1\right]}
\label{eps}
\ee
is the effective dielectric function of the structure air -- dielectric layer (thickness $d$, dielectric constant $\kappa$) -- 2DEG -- dielectric substrate (infinite thickness, dielectric constant $\kappa$). If $d=0$ (the thickness $d$ is typically very small, $d\lesssim 0.1$ $\mu$m), the dispersion relation (\ref{de})--(\ref{eps}) can be plotted as the dimensionless wave-vector $[\omega_p(q)/\omega]^2\propto q$ versus dimensionless magnetic field $\omega_c/\omega\propto B$; here $\omega_p^2(q)=2\pi n_se^2q/m^\ast\bar\kappa$ is the 2D plasmon frequency at $B=0$ and $\bar\kappa=(\kappa+1)/2$. This dependence is shown by the black solid curve in Fig. \ref{fig.4}. Other curves in this Figure show the $q$-vs-$B$ dependence at the finite dielectric cover-layer thickness $d$. At a finite $d$ the EMP spectrum additionally depends on the dimensionless parameter $2\pi n_se^2/m^{\ast}\bar \kappa d\omega^2$; for the density $n_s$ and the frequency $f$ we have chosen parameters of our experiment, see Figure \ref{fig.4} caption. The theoretical curves start (at $B=0$) from the value $[\omega_p(q)/\omega]^2=\eta_0\approx 1.217$ which determines the spectrum of edge plasmons in zero magnetic field, see Eq. (39) in Ref. \cite{Volkov88Eng}.

To compare now the theoretical curves of Fig. \ref{fig.4} with the experimentally found EMP-interference maxima in Fig. \ref{fig.3}(a) we introduce in (\ref{eps}) the length $L$, $q\to qL/L$, and consider Eqs. (\ref{de})--(\ref{eps}) as an implicit relation between the interference parameter $qL/2\pi$ and four dimensionless quantities
\be 
\frac{2\pi n_se^2}{m^{\ast}\kappa L\omega^2},\ \ 
\frac{2\pi n_se^2}{m^{\ast}\kappa d\omega^2},\ \ 
\frac{eB}{m^{\ast}c \omega},\ \ \textrm{and} \ \ \kappa.\label{par}
\ee 
The electron density $n_s$, the GaAs dielectric constant ($\kappa=12.8$) and the thickness of the dielectric layer above the 2DEG ($d=90$ nm) are known in Eq. (\ref{par}), therefore we can plot $qL/2\pi$ as a function of $B$ and fit these curves to the experimental points considering $L$ as a fitting parameter. 

Specifically, we proceed as follows. Consider, for example, the upper (magenta) curve in Figure \ref{fig.3}(a). It corresponds to the frequency $f=166$ GHz and has eight oscillation periods (nine maxima) in the interval $0.86\le B\le 1.92$ T. Plotting the theoretical curve we choose the length $L$ so that inside this interval there were exactly eight oscillation periods. This gives us both the integer indexes (from $n=13$ to $n=21$), unambiguously ascribed to each of the nine maxima of $R(B)$, and the length $L\approx 880$ $\mu$m, see Figure \ref{fig.3}(b). The accuracy of this method is quite high: a closer look at the Figure shows that the two thin magenta curves corresponding to 870 and 890 $\mu$m give a worse fit to the experimental data. 

Using the same method we fit three other sets of experimental data from Figure \ref{fig.3}(a). The lengths found for 148, 132 and 128 GHz are $L=870$, 860 and 840 $\mu$m, Figure \ref{fig.3}(b). Thus found values of $L$ slightly decrease with the frequency but the whole change of $L$ for $f$ lying between 166 and 128 GHz does not exceed 5 \%. Thus we conclude that the characteristic length which determines the EMP-related oscillations is much larger than the distance $L_{ab}$ in Figure \ref{fig.1} (inset) and is of the mm scale. This corresponds, approximately, to the distance between the points $c$ and $g$ in our experiment.

\begin{figure}[!h]
\includegraphics[width=\columnwidth]{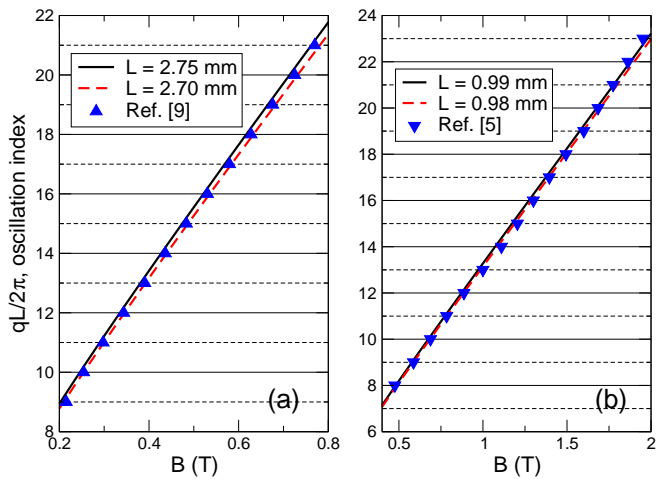}
\caption{\label{fig:stone}(Color online) The fit of the experimental data from (a) Fig. 1b of Ref. \cite{Stone07} (sample A, $f=37.5$ GHz, $L_{ab}=1$ mm, $\Delta R$ curve) and (b) Fig 1a of Ref. \cite{Kukushkin04a} ($f=53$ GHz, $L_{ab}=0.5$ mm).  } 
\end{figure}

Similarly, we have also analyzed the data of Refs. \cite{Stone07} and \cite{Kukushkin04a}. Figure \ref{fig:stone} shows the results of the fitting procedure for two selected experimental curves from these papers. The interference lengths $L$ found this way are about $2.7$ and $1$ mm respectively which is also larger than the corresponding distances $L_{ab}=1$ or $L_{ab}=0.5$ mm in Refs. \cite{Stone07} and \cite{Kukushkin04a}. Thus both in our work and in the previous publications \cite{Kukushkin04a,Stone07} the length $L$ extracted from the experimentally observed $B$-periodic oscillations is substantially larger than $L_{ab}$.

What exactly determines the interference length $L$ is difficult to ascertain because the real geometry of the 2DEG with attached contact pads is complicated in our experiment and unknown in Refs. \cite{Stone07,Kukushkin04a}. In all cases, however, the length $L$ is close to the average distance between the contact pads, counted if to move along the boundary of the 2DEG (or, in our work, between the contact pad and the QPC). This agrees with the simple physical picture of EMPs running along the edge of the 2DEG and generated at the boundaries of regions with \textit{substantially different} electron densities,\cite{Kukushkin04a} i.e., at the boundaries 2DEG -- metal contacts (the points $c$, $d$ in Figure \ref{fig.1}(b)).

One more interesting feature that we have observed in our experiment is the very large amplitude of $B$-periodic oscillations ($\Delta R/R \simeq 7$) which is substantially larger than in previous publications \cite{Kukushkin04a,Kukushkin05a,Stone07,Dorozhkin05} ($\Delta R/R \lesssim 1$). Actually the oscillations in our experiment look like a set of narrow resonances with a high quality factor. We attribute this very useful feature to the use of the bridged-gate QPC which was shown to substantially increase the local microwave field in the near contact area, see Ref. \cite{Levin15}.

To summarize, we have experimentally studied the $B$-periodic microwave induced magnetoresistance oscillations in a GaAs/AlGaAs quantum well samples. Comparing experimental results with the theory we have established that the EMP interference length responsible for the period of the magnetoresistance oscillations is presumably determined by the distance between real metallic contacts (contact pads) measured along the boundary of the 2DEG. This partly explains the contradiction between the earlier published results \cite{Kukushkin04a,Stone07}, but in order to reliably establish which length or lengths are ultimately responsible for the formation of EMP-related oscillations further experimental studies on samples with possibly simple geometry of the 2DEG and contact pads are needed, together with the corresponding theoretical modeling. We have also demonstrated that using the bridged-gate QPC one can substantially increase the amplitude of the $B$-periodic oscillations and hence the sensitivity of the microwave detectors and spectrometers based on the discussed phenomenon. 

We thank N. A. Savostianova for assistance and O. E. Raichev for helpful discussions. The financial support of this work by FAPESP, CNPq (Brazilian agencies), RFBI (No 17-02-00384) and RAS (No 16-116021510194-3) is acknowledged.

%


\end{document}